\RequirePackage[columnwise,running]{lineno}
\documentclass[aps,pra,
amssymb,nofootinbib,twocolumn,longbibliography]{revtex4}

\usepackage{graphicx}
\usepackage{dcolumn}
\usepackage{bm}
\usepackage{bbold}
\usepackage{lineno}
\usepackage{amsmath}
\usepackage{verbatim}
\usepackage[applemac]{inputenc}
\usepackage{nameref}
\usepackage{varioref}
\usepackage{hyperref}
\usepackage{amssymb}
\usepackage{bbm, bm, amsbsy, amsthm, amsfonts, amsmath, dsfont, graphicx, epsfig, epstopdf, dsfont, multibib, color,relsize}
\usepackage{cleveref}

\usepackage{color}

\newcommand{\bra}[1]{{\left\langle{#1}\right\vert}}
\newcommand{\ket}[1]{{\left\vert{#1}\right\rangle}}

\newcommand{\avg}[1]{\langle{#1}\rangle}

\let\oldequation\equation
\let\oldendequation\endequation

\renewenvironment{equation}
  {\linenomathNonumbers\oldequation}
  {\oldendequation\endlinenomath}

\begin{document}

\title{Experimental quantum speed-up in reinforcement learning agents}

\author{V. Saggio$^{1*}$}

\author{B. E. Asenbeck$^1$}

\author{A. Hamann$^{2}$}

\author{T. Str\"{o}mberg$^{1}$}

\author{P. Schiansky$^{1}$}

\author{V. Dunjko$^{3}$}

\author{N. Friis$^{4}$}

\author{N. C. Harris$^{5,6}$}

\author{M. Hochberg$^{7}$}

\author{D. Englund$^{6}$}

\author{S. W\"{o}lk$^{2,8}$}

\author{H. J. Briegel$^{2,9}$}

\author{P. Walther$^{1,10}$}

\affiliation{\vspace{0.4cm}
$^1$University of Vienna, Faculty of Physics, Vienna Center for Quantum Science and Technology (VCQ), Boltzmanngasse 5, Vienna A-1090, Austria}

\affiliation{\vspace{0.1mm}
$^2$Institut f\"ur Theoretische Physik, Universit\"at Innsbruck, Technikerstra{\ss}e 21a, 6020 Innsbruck, Austria}

\affiliation{\vspace{0.1mm}
$^3$LIACS,  Leiden  University,  Niels  Bohrweg  1,  2333  CA  Leiden,  Netherlands}

\affiliation{\vspace{0.1mm}
$^4$Institute for Quantum Optics and Quantum Information - IQOQI Vienna, Austrian Academy of Sciences, Boltzmanngasse 3, A-1090 Vienna, Austria}

\affiliation{\vspace{0.1mm}
$^5$Lightmatter, 60 State Street, Floor 10, Boston, MA, 02130, USA}

\affiliation{\vspace{0.1mm}
$^6$Research Laboratory of Electronics, Massachusetts Institute of Technology, Cambridge, MA 02139, USA}

\affiliation{\vspace{0.1mm}
$^7$NOKIA Corporation, New York, NY, USA}

\affiliation{\vspace{0.1mm}
$^8$Deutsches Zentrum f{\"u}r Luft- und Raumfahrt e.V. (DLR), Institut f{\"u}r Quantentechnologien, S{\"o}flingerstr. 100, 89077 Ulm, Germany}

\affiliation{\vspace{0.1mm}
$^9$Fachbereich Philosophie, Universit{\"a}t Konstanz, Fach 17, 78457 Konstanz, Germany}

\affiliation{\vspace{0.1mm}
$^{10}$Christian Doppler Laboratory for Photonic Quantum Computer, Faculty of Physics, University of Vienna, Vienna, Austria.}

\begin{linenomath*}
\begin{abstract}
\vspace{-1mm}
Increasing demand for algorithms that can learn quickly and efficiently has led to a surge of development within the field of artificial intelligence (AI). An important paradigm within AI is reinforcement learning (RL), where agents interact with environments by exchanging signals via a communication channel. Agents can learn by updating their behaviour based on obtained feedback. The crucial question for practical applications is how fast agents can learn to respond correctly. An essential figure of merit is therefore the learning time. While various works have made use of quantum mechanics to speed up the agent's decision-making process, a reduction in learning time has not been demonstrated yet. Here we present a RL experiment where the learning of an agent is boosted by utilizing a quantum communication channel with the environment. We further show that the combination with classical communication enables the evaluation of such an improvement, and additionally allows for optimal control of the learning progress. This novel scenario is therefore demonstrated by considering hybrid agents, that alternate between rounds of quantum and classical communication. We implement this learning protocol on a compact and fully tunable integrated nanophotonic processor. The device interfaces with telecom-wavelength photons and features a fast active feedback mechanism, allowing us to demonstrate the agent’s systematic quantum advantage in a setup that could be readily integrated within future large-scale quantum communication networks.
\end{abstract}
\end{linenomath*}
\maketitle

\vspace{-2mm}
\section*{INTRODUCTION}
\vspace*{-2mm}
Rapid advances in the field of machine learning (ML) and in general artificial intelligence (AI) have been paving the way towards intelligent algorithms and automation. An important paradigm within AI is reinforcement learning (RL), where agents interact with an environment and `learning' is facilitated by feedback exchanges~\cite{sutton}. 
RL has applications in many sectors, from robotics (e.g. robot control~\cite{johannink}, text and speech recognition~\cite{tjandra}) to the healthcare domain (e.g. finding optimal treatment policies~\cite{komorowski}) to simulation of brain-like computing~\cite{thakur} and neural networks~\cite{shen,steinbrecher}. RL is also at the heart of the celebrated AlphaGo algorithm~\cite{silver}, able to beat even the most skilled human players at the game of Go.

At the same time, thanks to the capability of quantum physics to outperform classical algorithms, the development of quantum technologies has been experiencing remarkable progress~\cite{arute}.  
Quantum physics offers improved performance in a variety of applications, from quantum-enhanced sensing~\cite{giovannetti} to secure communication~\cite{sergienko} and quantum information processing~\cite{monroe}. 
Quantum mechanics also inspires new advantageous algorithms in ML, and in particular RL~\cite{dong,dunjko}. 
In fact, RL has been successfully implemented to aid in problems encountered in quantum information processing itself, e.g. decoding of errors~\cite{baireuther, breuckmann, chamberland}, quantum feedback~\cite{fosel}, adaptive code-design~\cite{poulsen}, 
and even the design of new quantum experiments~\cite{krenn,melnikov}. Conversely, quantum technology has been employed to enable a quadratically faster decision-making process for RL agents 
via the quantization of their internal hardware~\cite{paparo, dunjko2, jerbi,sriarunothai}.

In all of these RL applications, the interaction between the agent and the environment is performed via entirely classical communication. 
Here we consider a novel RL setting where the agent and the environment can also communicate via a quantum channel~\cite{dunjko3}. In particular, we introduce a hybrid agent that enables quantum as well as classical information transfer, and thus combines quantum amplitude amplification with a classical update policy via a feedback loop. Within such a scenario, it is possible for the first time to quantify and achieve a quantum speed-up in the agent's learning time with respect to RL schemes based on classical communication only. 
\begin{figure}[ht!]
\centering
\includegraphics[width=0.482\textwidth]{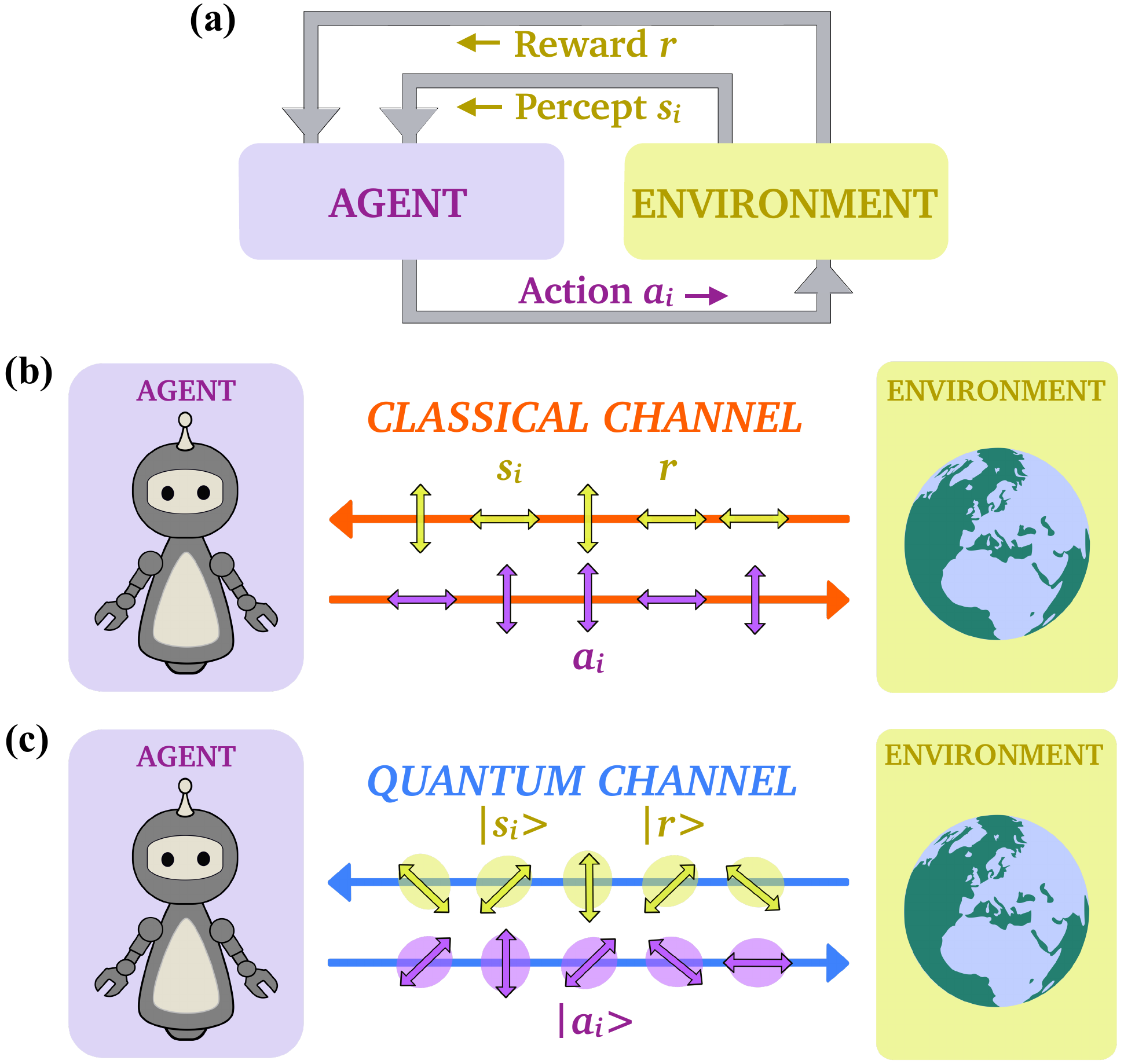}
\caption{\textbf{Schematic of a learning agent.} \textbf{(a)} An agent interacts with an environment by receiving perceptual input $s_i$ and outputting consequent actions $a_i$. In the case of the correct $a_i$ being performed, the environment issues a reward $r$ that the agent uses to enhance its performance in the next round of interaction. \textbf{(b)} Agent and environment interacting via a classical channel, where communication is only possible via a fixed preferred basis (e.g. `vertical' or `horizontal'). \textbf{(c)} Agent and environment interacting via a quantum channel, where arbitrary superposition states are exchanged. }
\label{rl}
\end{figure}
The learning time is defined as the average number of interactions until the agent receives a reward (`accomplishes a specific task'). Such improvement represents a remarkable advantage over previously implemented protocols, and may prove crucial in the development of increasingly complex learning devices~\cite{thakur,shen,steinbrecher}.

We demonstrate this protocol using a fully-programmable nanophotonic processor interfaced with photons at telecom wavelengths. The setup enables the implementation of an active feedback mechanism, thus proving suitable for demonstrations of RL algorithms.  
Moreover, such a photonic platform holds the potential of being fully interfaceable with future quantum communication networks thanks to the photons' telecom wavelengths. In fact, a long-standing goal of the current development of quantum communication technologies lies in establishing a form of `quantum internet’~\cite{kimble,cacciapuoti} --- a highly interconnected network able to distribute and manipulate complex quantum states via fibres and optical links (ground-based or even via satellites). We therefore envisage AI and RL to play an important role in future large-scale quantum communication networks, including a potential quantum internet, much in the same way that AI forms integral parts of the internet today. 
Our results thus additionally demonstrate the feasibility of integrating quantum mechanical RL speed-ups in future complex quantum networks.

\vspace*{-2mm}
\section*{QUANTUM ENHANCEMENT IN REINFORCEMENT LEARNING}
\label{secQRL}
\vspace*{-2mm}

The conceptual idea of RL is shown in Fig.~\ref{rl}(a).
Here an agent (e.g. physical or web-based) interacts with an environment by receiving perceptual input, called `percepts' $s_i$, and outputting specific `actions' $a_i$ accordingly. `Rewards' $r$ issued by the environment for correct combinations of percepts and actions incentivize agents to improve their decision-making, and thus to learn~\cite{sutton}.   

Although RL has already been shown amenable to quantum enhancements, in practical demonstrations the interaction between agents and environments has so far been restricted exclusively to classical communication, meaning that agents are limited to exchanges in a fixed preferred basis, e.g. `vertical' or `horizontal' as shown in Fig.~\ref{rl}(b). In general, it has been shown that granting an agent access to quantum hardware (while still restricting it to classical exchanges with the environment) does not reduce the average number of interactions that the agent needs to accomplish its task, although it allows it to output actions quadratically faster~\cite{paparo,sriarunothai}. 
To achieve a reduction in learning times, and thus in sample complexity, fully quantum-mechanical interactions between agent and environment must be considered.

\begin{figure*}[ht]
\centering
\includegraphics[width=6.8in]{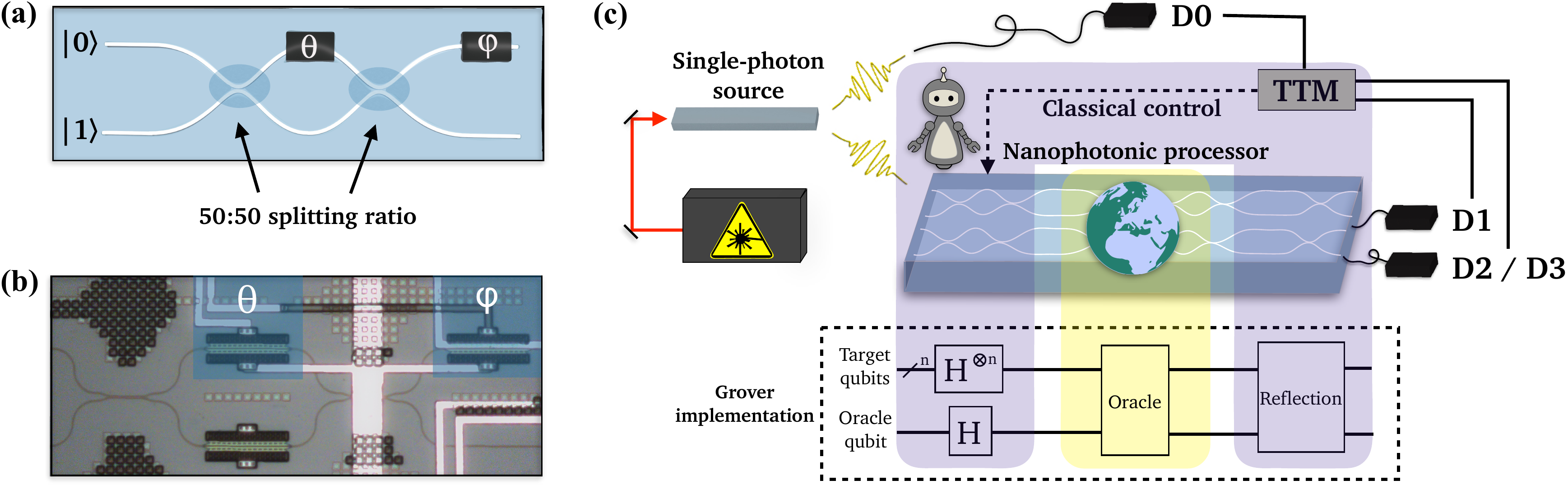}
\caption{\textbf{Experimental setup.} \textbf{(a)} Single programmable unit consisting of a Mach-Zehnder interferometer (MZI) equipped with two fully tunable phase shifters $\theta$ and $\phi$. \textbf{(b)} Real-life picture of a single MZI in the processor. The third phase shifter in the bottom arm of the interferometer is not used. \textbf{(c)} Overview of the setup. A single-photon source generates single-photon pairs at telecom wavelength. One photon is sent straight to a single-photon detector D0, while the other one is coupled into the processor and undergoes the desired computation. It is then detected, in coincidence with the photon in D0, either in detector D1 or D2/D3 after the agent plays the classical/quantum strategy (refer to Fig.~\ref{circuits} for more details). The coincidence events are recorded with a custom-made Time Tagging Module (TTM). Different areas of the processor are assigned to either the agent or the environment. The former one also comprises a classical control updating its policy. They alternately play a Grover-like search to look for the rewarded sequence of actions.}
\label{chip}
\end{figure*}

Our goal is to experimentally demonstrate such a quantum-enhanced RL scenario. 
We start with considering an agent and environment with access to internal quantum (as well as classical) hardware 
that can interact by coherently exchanging quantum states $\ket{a_{i}}$, $\ket{s_{i}}$, and $\ket{r}$, representing actions $a_{i}$, percepts $s_{i}$, and rewards $r$, respectively, via a joint quantum channel, as illustrated in Fig.~\ref{rl}(c). 
In this case, communication is no longer limited to a fixed preferred basis, but allows for an exchange of arbitrary superpositions.
Agents react to (sequences of) percepts $\ket{s_{i-1}}$ with (sequences of) actions $\ket{a_i}$ according to a policy $\pi(a_i|s_{i-1})$ that is updated during the learning process. 
In particular, this quantum framework includes so-called deterministic strictly epochal (DSE) learning scenarios, which we focus on. Here, `epochs' consist of strings of percepts $\vec{s}=(s_0,\cdots,s_{L-1})$ and actions $\vec{a}=(a_1,\cdots,a_L)$ of fixed length $L$, and a final reward $r$, and both $\vec{s}=\vec{s}(\vec{a})$ and $r=r(\vec{a})$ are completely determined by $\vec{a}$ (see Methods, Sec.~\ref{Append1}). Many interesting environments are epochal, e.g. in applications of RL to quantum physics~\cite{denil,melnikov,bukov,poulsen2} or popular problems such as playing Go~\cite{silver}. 
A non-trivial feature of the DSE scenario is that the effective behaviour of the environment 
can be modelled via a unitary $U_{E}$ on the action and reward quantum registers (indicated with subscripts $\mathrm{A}$ and $\mathrm{R}$, respectively):
\begin{equation}
    U_E\ket{\vec{a}}_{\mathrm{A}}\ket{0}_{\mathrm{R}}=\left\lbrace \begin{array}{cc}\ket{\vec{a}}_{\mathrm{A}}\ket{1}_{\mathrm{R}} & \text{ if } r(\vec{a})>0\\ \ket{\vec{a}}_{\mathrm{A}}\ket{0}_{\mathrm{R}} & \text{else}
\end{array}\right.,
\label{equU}
\end{equation}
which identifies rewarded sequences of actions. The quantum-enhanced agent can use $U_E$ to perform a quantum search for rewarded action sequences and teach the found sequences to a classical agent~\cite{dunjko3}. We push this idea  further and grant the quantum and the classical part of the agent complete access to each other. In this way, it is possible to create a hybrid agent with a feedback loop between quantum search and classical policy update. This approach allows us to quantify the speed-up in learning time, which is not possible in the general setting discussed in~\cite{dunjko3} (see Methods, Sec. \ref{Append1}).

It is known that near-term quantum devices are generally subject to limited coherence times, restricting the number of quantum gates that can be performed before essential quantum features are lost. However, already a few or even a single coherent query to $U_{E}$ are enough to construct a quantum-enhanced RL framework that can outperform a classical one. 

In a classical epoch, the agent prepares the state $\ket{\vec{a}}_{\mathrm{A}}\ket{0}_{\mathrm{R}}$ where $\vec{a}$ is determined by directly sampling from a classical probability distribution $p(\vec{a})$. From classically interacting with the environment, the agent determines the corresponding reward $r(\vec{a})$ and percepts $\vec{s}(\vec{a})$, 
and uses the obtained information to update its policy $\pi$, thus learning. 
 
In a quantum epoch, the agent prepares the state $\ket{\psi}_{\mathrm{A}}\ket{-}_{\mathrm{R}}$, with $\ket{\psi}= \sum_{\vec{a}}\sqrt{p(\vec{a})}\,\ket{\vec{a}}$ being determined according to $p(\vec{a})$ dictated by its current classical policy $\pi$, and $\ket{-}=(\ket{0}-\ket{1})/\sqrt{2}$. Both states are sent to the environment, which acts on them via $U_E$  inducing a phase of $-1$ for all rewarded sequences of actions. The states are sent back to the agent which performs a reflection $U_{Ref}=(2\ket{\psi} \bra{\psi}_{\mathrm{A}}-\mathds{1}_{\mathrm{A}})$ over the initial superposition state. This leads to an amplitude amplification of rewarded action sequences similar to the Grover's algorithm~\cite{grover}. Parametrising the classical probability to find a reward via $\sin^2(\xi)$ with $\xi \in [0,2 \pi]$, the steps above followed by a final measurement on the action register result with an increased probability $\sin^2(3\xi)$ in a rewarded action sequence. However, this procedure does not reveal the reward or the corresponding sequence of percepts. These can be determined only during classical test epochs, where the measured sequence of actions is used as input state.

The quantum-enhanced agent plays in a hybrid way by alternating between these quantum and classical test epochs. 
Such agents find rewarded sequences of actions faster, and hence usually  learn faster than entirely classical agents. The learning speed-up manifests in a reduced average quantum learning time $\avg{T}_Q$, that is, the average number of epochs necessary to achieve a predefined winning probability $Q_L$. In general, a quadratic improvement in the learning time given by $\avg{T}_Q\leq \alpha \sqrt{J \avg{T}_C}$ for the hybrid agent can be achieved if the maximal number of coherent interactions between the agent and the environment scales with the problem size. Here, $J$ quantifies the number of rewards the agent needs to find in order to learn and $\avg{T}_C$ the average classical learning time.  However, already a single (coherent) query to $U_E$ is enough to improve the learning time albeit by a constant factor, as demonstrated in the Methods, Sec.~\ref{Append1}.

\vspace*{-2mm}
\section*{EXPERIMENTAL IMPLEMENTATION}\label{sectionImp}
\vspace*{-2mm}
Quantized RL protocols can be compactly realized using state-of-the-art photonic technology~\cite{flamini}. 
In particular the path towards miniaturization of photonic platforms holds the advantage of providing scalable architectures where many elementary components can be accommodated on small devices. 
Here we use a programmable nanophotonic processor with dimensions of $4.9$ x $2.4$ mm, comprising $26$ waveguides fabricated to form $88$ Mach-Zehnder interferometers (MZIs) in a trapezoidal configuration. A quantum gate is implemented by a MZI equipped with two thermo-optic phase shifters, one internal to allow for a scan of the output distribution over $\theta \in [0, 2 \pi ]$ and one external dictating the relative phase $\phi \in [0, 2 \pi ]$ between the two output modes, as shown in Fig.~\ref{chip}(a). This makes each MZI act as a fully-tunable beam splitter and allows for coherent implementation of sequences of quantum gates. Information is spatially encoded onto two orthogonal modes $\ket{0}=(1,0)^\mathrm{T}$ and $\ket{1}=(0,1)^\mathrm{T}$, which constitute the computational basis. Fig.~\ref{chip}(b) shows how a MZI looks like in our integrated device. 

Pairs of single photons are generated in the telecom wavelength band from a single-photon source pumped by laser light at 789.5 nm. One photon is coupled into one waveguide in the processor to perform a particular computation, while the other one is sent straight to a single-photon detector D0 for heralding. The detectors are superconducting nanowires that combine extremely high efficiency (up to $\sim$90\%) with a very short dead time ($<$ 100 ns), thus proving the best candidates for fast feedback at telecom wavelengths. Detection events recorded in D0 and at the output of the processor falling into a temporal window of 1.3 ns are registered with a time tagging module (TTM) as coincidence events (see Methods, Sec.~\ref{Append2} for more details). Different areas of the processor are assigned to the agent and the environment, who alternately perform the aforementioned steps of the Grover-like amplitude amplification. The agent is further equipped with a classical control mechanism that updates its learning policy. All of this can be seen in Fig.~\ref{chip}(c).

\begin{figure}[ht]
\centering
\includegraphics[width=3.5in]{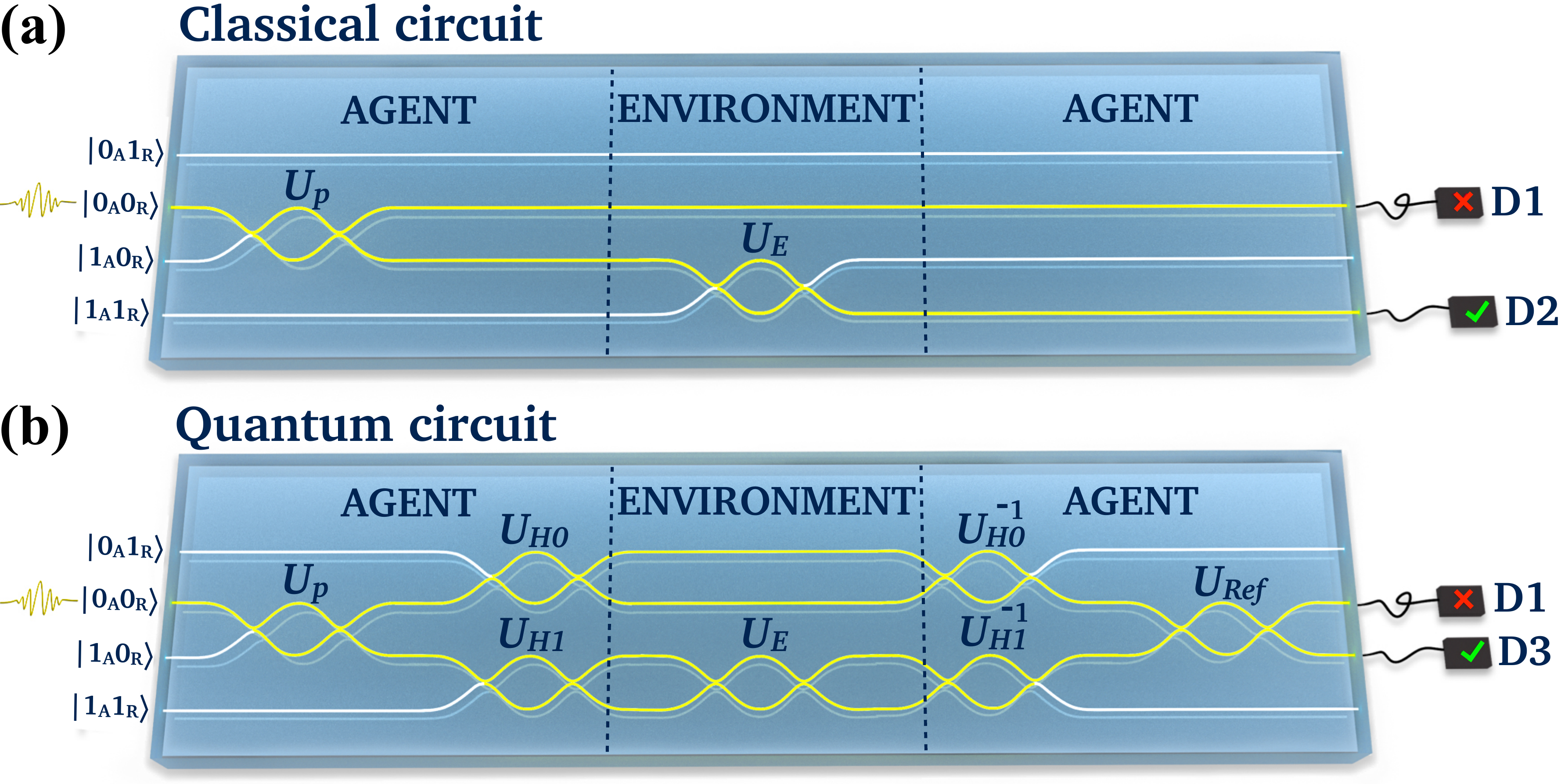}
\caption{\textbf{Circuit implementation.} One photon is coupled into the $\ket{0_{\mathrm{A}}0_{\mathrm{R}}}$ spatial mode and undergoes different unitaries depending on whether a classical \textbf{(a)} or a quantum \textbf{(b)} epoch is implemented. The waveguides highlighted in yellow show the photon's possible paths. Identity gates are represented with straight waveguides.}
\label{circuits}
\end{figure}

In our experiment, we represent all possible sequences of actions by a  single qubit ($\ket{1}_\mathrm{A}\widehat{=}$ rewarded, $\ket{0}_\mathrm{A}\widehat{=}$ not rewarded) and use another qubit to encode the reward ($\ket{1}_\mathrm{R}$, $\ket{0}_\mathrm{R}$). This results in a four-level system where each level is a waveguide path in our processor, as shown in Fig.~\ref{circuits} (which illustrates only the part of the processor needed for our computation).

The probability for the agent to choose a correct sequence of actions is initially set to $\sin^2(\xi)=\varepsilon=0.01$, representing, for instance, a single rewarded sequence of actions out of 100. After a single photon is coupled into the mode $\ket{0_\mathrm{A} 0_\mathrm{R}}$, the agent puts it into the superposition state $\ket{\psi}_{\mathrm{A}}=(\cos{\xi}\ket{0}_\mathrm{A}+ \sin{\xi}\ket{1}_\mathrm{A}) \ket{0}_\mathrm{R}$ via the unitary $U_p$. Next, it can decide to play classically or quantum-mechanically. 

\begin{figure*}[ht]
\centering
\fontsize{9pt}{10pt}\selectfont
\def\svgwidth{7.0in}
\includegraphics[width=7.1in]{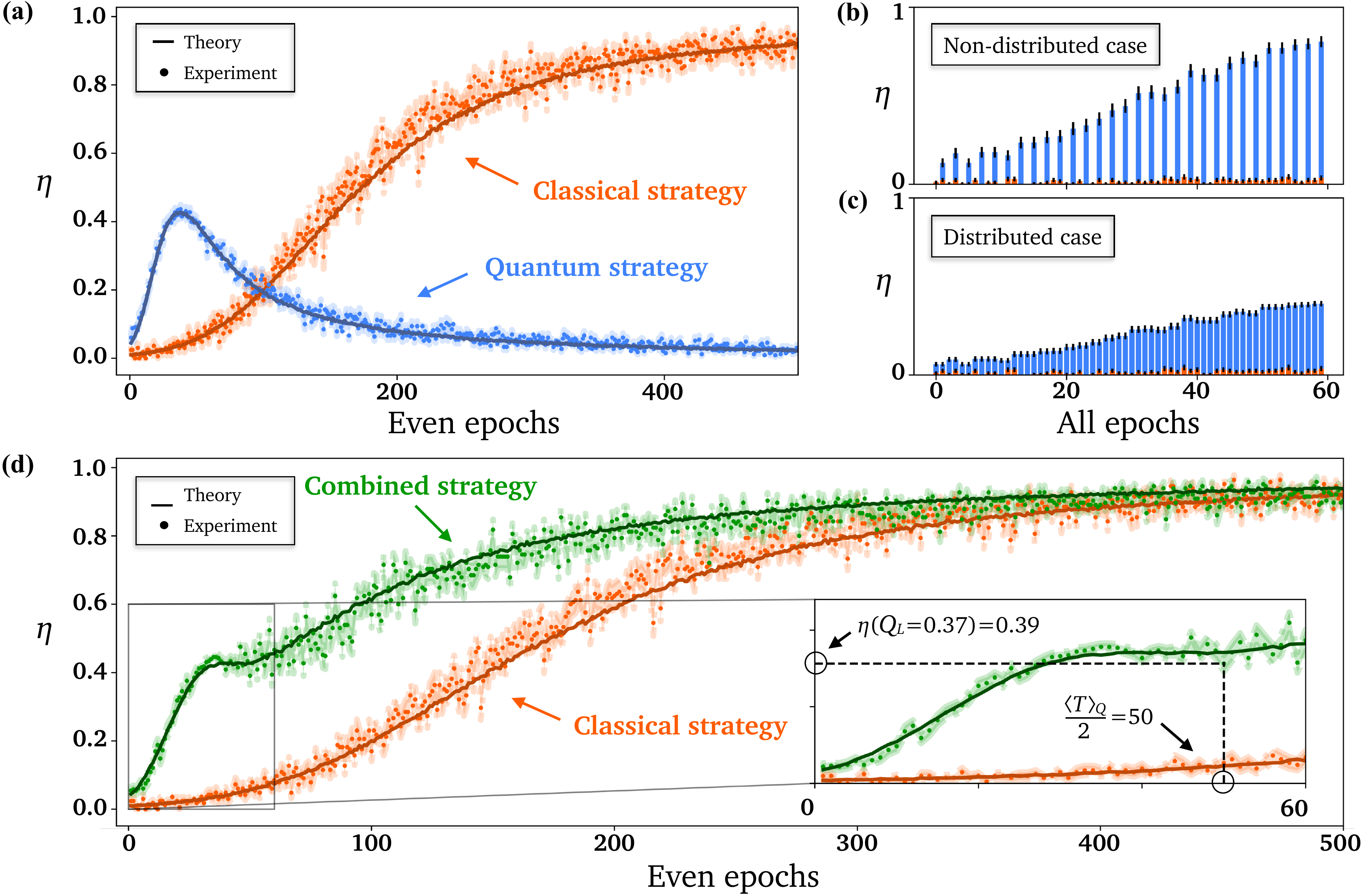}
\caption{\textbf{Behaviour of the average reward $\eta$ for different learning strategies.} The solid line represents the theoretical data simulated with $n=10,000$ agents, while the dots represent the experimental data measured with $n=165$ agents. The shaded regions indicate the errors associated to each single data point. \textbf{(a)} $\eta$ of agents playing a quantum (blue) or classical (orange) strategy. \textbf{(b)} $\eta$ accounting, in the quantum case, for rewards obtained only every second epoch, compared to \textbf{(c)} the case where the reward is distributed over the two epochs an agent needs to acquire it. 
\textbf{(d)} Comparison between the classical and combined quantum-classical case, where an advantage over the classical case is visible. Here, the agents stop the quantum strategy at their best performance (at $\varepsilon=0.396$) and continue playing classically. The inset shows the point at which the agent playing the quantum strategy reaches the predefined winning probability $Q_L=0.37$, after $\avg{T}_{Q}=100$ epochs.}
\label{quantum}
\end{figure*}

\textit{Classical strategy.} The environment flips the reward qubit only if the action qubit is in the correct state via the 
unitary $U_{E}$, as shown in Fig.~\ref{circuits}(a). 
Next, the photon is coupled out and detected in either D1 or D2 with probability $\cos^2(\xi)$ and $\sin^2(\xi)$, respectively. Only if D2 is triggered (i.e. the agent has been rewarded), a classical feedback mechanism updates the policy $\pi$, and thus $p(\vec{a})$ and $\varepsilon$, via rules given by the projective simulation framework (see Methods, Sec.~\ref{Append1}). 

\textit{Quantum strategy.} Here, the theoretical framework described above is exploited to speed up the learning process.  
As illustrated in Fig.~\ref{circuits}(b), after the reward qubit is rotated to $\ket{-}_{\mathrm{R}}$ 
via the unitaries $U_{H0}$ and $U_{H1}$, the environment acts as an oracle via $U_{E}$. 
Consecutively, the agent reverses the effect of $U_{H0}$ and $U_{H1}$, and then it performs the reflection $U_{Ref}$. 
Next, a measurement in the computational basis of the action register is performed, leading to detection of a rewarded sequence of actions in D3 with increased probability $\sin^2(3 \xi)$.  The classical test epoch needed to determine the reward is not experimentally implemented for practical reasons, since the circuit visibility is bigger than 0.99.

In general, any Grover-like algorithm faces a drop in the amplitude amplification after the optimal point is reached. Keeping in mind that each agent will reach this optimal point at different epochs, one can identify the probability $\varepsilon=0.396$ (see Methods, Sec.~\ref{Append1}) until which it is beneficial for all agents to use a quantum strategy, as on average they will observe more rewards than with the classical strategy. As soon as this probability is surpassed, it is advantageous for the agents to switch to an entirely classical strategy. This combined quantum-classical strategy thus avoids the winning probability drop without introducing any additional overheads in terms of experimental resources.

\vspace*{-2mm}
\section*{RESULTS}
\vspace*{-2mm}

Here, we show the experimental comparison between quantum and classical strategies. 

At the end of each classical epoch, we record outcomes $1$ and $0$ for the rewarded and not rewarded behaviour, obtaining a binary sequence whose length equals the number of played epochs for the classical learning strategy, and half of the number of epochs for the quantum strategy. 
To enable a fair comparison between the quantum and classical scenarios, the reward in the quantum case is distributed --- i.e. averaged --- over the two epochs needed to determine a test sequence of actions $\vec{a}$ and its reward.
The reward is then averaged over many different agents playing independently of one another. Fig.~\ref{quantum} shows the average reward $\eta$ for the different learning strategies.

The theoretical data is simulated for $ n = 10,000$ agents and the experimental data obtained from $n=165$. Fig.~\ref{quantum}(a) visualizes the quantum improvement originating from the use of amplitude amplification in comparison with a purely classical strategy. 
For completeness, we also show in Fig.~\ref{quantum}(b) and (c) the comparison between not distributing and distributing the reward over two epochs in the quantum case. 

As soon as $\varepsilon=0.396$ is reached, 
$\eta$ starts decreasing. 
Our setup allows the agents to always choose the favorable strategy by switching from quantum to classical as soon as the second becomes more advantageous. In this way, such combined strategy always outperforms the purely classical scenario, as shown in Fig.~\ref{quantum}(d).

We define $Q_L = 0.37$ as the winning probability. Note however that any probability below $\varepsilon=0.396$ can be defined as the winning probability $Q_L$.
The learning time $\avg{T}$ for $Q_L$ decreases from $\avg{T}_C=270$ in the classical case to $\avg{T}_{Q}=100$ in the combined quantum-classical case. This implies a reduction of $63\%$, which fits well to the theoretical values $T_C^{theory}=293$ and $T_{Q}^{theory}=97$, taking into account small experimental imperfections.

As the combined configuration prevents the average reward from dropping, 
it is particularly relevant for real-life applications when a Grover search (with its intrinsic overshooting drawback) is implemented.
An average reward that saturates at a high niveau without a subsequent drop can also be achieved by employing other algorithms like the fixed-point algorithm~\cite{yoder}. However, they show a less favorable speed-up than Grover-like amplitude amplification, especially considering the limited size of our integrated processor.

In general, an agent can experience a quadratic speed-up in its learning time if it can perform an arbitrary number of coherent Grover iterations~\cite{hamann} even if the number of actual rewarded sequences is unknown~\cite{boyer}.

\vspace*{-4mm}
\section*{CONCLUSIONS}
\vspace*{-3mm}

We have demonstrated a novel RL protocol where information is alternately communicated via a quantum and a classical channel. This makes it possible to evaluate the agent's performance resulting in a speed-up in the learning time, and gain optimal control of the learning process. Learning agents relying on purely classical communication are therefore outperformed. At the same time, emerging photonic circuit technology provides the advantages of compactness, full tunability and low-loss communication, thus proving suitable for RL algorithms where active feedback mechanisms, even over long distances, need to be implemented.

We highlight that although photonic architectures prove particularly suitable for these types of learning algorithms, the presented demonstration is based on a novel theoretical background that is general and applicable to any quantum platform. Moreover, as the field of integrated optics is moving towards the fabrication of increasingly large devices, this demonstration could be extended to more complex quantum circuits allowing for processing of high-dimensional states, and thus paving the way for achieving superior performance in increasingly complex learning devices.\\

\noindent{$^*$Corresponding author: valeria.saggio@univie.ac.at.}

\vspace{1cm}

\noindent{\textbf{Acknowledgments:} The authors thank Lee A. Rozema, Irati Alonso Calafell and Philipp Jenke for help with the detectors.
 A.H. acknowledges support from the Austrian Science Fund (FWF) through the project P 30937-N27. V.D. acknowledges support from the Dutch Research Council (NWO/ OCW), as part of the Quantum Software Consortium programme (project number 024.003.037). N.F. acknowledges support from the Austrian Science Fund (FWF) through the project P 31339-N27. H.J.B. acknowledges support from the Austrian Science Fund (FWF) through SFB BeyondC F7102, the Ministerium für Wissenschaft, Forschung, und Kunst Baden-Württemberg (Az. 33-7533-30-10/41/1) and the Volkswagen Foundation (Az. 97721). P.W. acknowledges support from the research platform TURIS, the European Commission through ErBeStA (no. 800942), HiPhoP (no. 731473), UNIQORN (no. 820474), EPIQUS (no. 899368), and AppQInfo (no. 956071), from the Austrian Science Fund (FWF) through CoQuS (W1210-N25), BeyondC (F 7113) and Research Group (FG 5), and Red Bull GmbH. The MIT portion of the work was supported in part by AFOSR award FA9550-16-1-0391 \textbf{and NTT Research}.
\textbf{Author contributions:} 
V.S. and B.E.A. implemented the experiment and performed data analysis. A.H., V.D., N.F., S.W. and H.J.B. developed the theoretical idea. T.S. and P.S. provided help with the experimental implementation. N.C.H., M.H., and D.E. designed the nanophotonic processor. V.S., S.W., and P.W. supervised the project. All the authors contributed to writing the paper. \textbf{Competing interests:} The authors declare they have no competing interests. \textbf{Data and material availability:} Data related to this work may be available upon request.}

\clearpage

\hypertarget{sec:appendix}
\appendix
\section*{METHODS}
\renewcommand{\thesubsection}{\Roman{subsection}}
\renewcommand{\thesubsubsection}{\Roman{subsection}.\arabic{subsubsection}}
\setcounter{equation}{0}
\renewcommand\theequation{A.\arabic{equation}}
\subsection{Quantum-enhanced 
learning agents}\label{Append1}

In this section, we develop an explicit method for combining a classical reinforcement learning agent with quantum amplitude amplification. Our approach for such hybrid agents goes beyond the ideas of Ref.~\cite{dunjko3} by introducing a feedback loop between classical policy update and quantum amplitude amplification. The developed model allows us to determine achievable improvements in sample complexity, and thus learning time. In addition, the final policy of our agent has similar properties as the policy of the underlying classical agent, leading to a comparable behavior as discussed in more detail in \cite{hamann} (a paper dedicated to discuss the theoretical background of the hybrid agent presented here more specifically).    

In the following, we concentrate on simple deterministic, strictly epochal (DSE) environments. That is, the interaction between the agent and the environment is structured into epochs where each epoch starts with the same percept $s_0$, and at each time step $i$ an action-percept pair $(a_i,s_i)$ is exchanged. At the end of each epoch, after $L$ action-percept pairs are communicated, a reward $r\in \lbrace 0,1 \rbrace$ is given to the agent. The rules of the game are deterministic and time independent, such that performing a specific action $a_i$ after receiving a percept $s_{i-1}$ always leads to the same following percept $s_i$.

The behaviour of an agent is determined by its policy described by the probability $\pi(a_i|s_{i-1})$ to 
perform the action $a_i$ given the percept $s_{i-1}$. In deterministic settings, the percept $s_i$ is 
completely determined by all previous performed actions $a_1, \cdots, a_i$ such that $\pi(a_i|s_{i-1})=\pi(a_i|a_1,\cdots, a_{i-1})$. Thus, the behaviour of the agent within one epoch is described by sequences of actions $\vec{a}=(a_1,\cdots,a_L)$ and their corresponding probabilities
\begin{eqnarray}
p(\vec{a})=\prod\limits_{i=1}^L \pi(a_i|a_1,\cdots,a_{i-1}).
\end{eqnarray}
The learning agent implemented here uses a policy based on projective simulation~\cite{briegel}, where each sequence of actions $\vec{a}$ is associated with a weight factor $h(\vec{a})$ initialized to $h=1$. Its policy is defined via the probability distribution
\begin{eqnarray}
p(\vec{a})&=&\frac{h(\vec{a})}{\sum\limits_{\vec{a}'}h(\vec{a}')}.
\end{eqnarray}
If the agent has played the sequence $\vec{a}$, it updates the corresponding weight factor via
\begin{eqnarray}
h(\vec{a})&\rightarrow &h(\vec{a})+\lambda r(\vec{a}),
\label{update}
\end{eqnarray}
where $\lambda=2$ in our experiment. In general, the update method for quantum-enhanced agents is not limited to projective simulation and can be used to enhance any classical learning scenario, provided that $p(\vec{a})$ exists and that the update rule is solely based on the observed rewards. 

We generalize the given learning problem to the quantum domain by encoding different sequences of actions $\vec{a}$ into orthogonal quantum states $\ket{\vec{a}}$ defining our computational basis. In addition, we create a fair unitary oracular variant $\tilde{U}_E$ of the environment~\cite{dunjko3}, whose behaviour is described by
\begin{eqnarray}
\tilde{U}_E \ket{\vec{a}}=\left\lbrace \begin{array}{cc}\ket{\vec{a}}& \text{ if } r(\vec{a})=0\\
-\ket{\vec{a}}& \text{ else }\end{array}\right. .
\end{eqnarray}
The unitary oracle $\tilde{U}_E$ can be used to perform, for instance, a Grover search or amplitude amplification for rewarded sequences of actions by performing Grover iterations
\begin{equation}
U_G=\Big( 2\ket{\psi}\bra{\psi}-\mathds{1}\Big)\tilde{U}_E
\end{equation}
on an initial state $\ket{\psi}$. A quantum-enhanced agent with access to $\tilde{U}_E$ can thus find rewarded sequences of actions faster than a corresponding classical agent, defined by the same initial policy $\pi(a_i|s_{i-1})$ and update rules, without access to $\tilde{U}_E$.

In general, the optimal number $k$ of Grover iterations $U_G^k\ket{\psi}$ depends on the winning probability
\begin{eqnarray}
q=\sum\limits_{\lbrace \vec{a}|r(\vec{a})\neq 0\rbrace} p(\vec{a})\label{eq:win_prob}
\end{eqnarray}
via $k\sim 1/\sqrt{q}$~\cite{grover}. In the following, we assume that $q$ is known at least to a good approximation. This is for instance possible if the number of winning sequences $\ket{\vec{a}}$ is known. However, a similar quantum-enhanced learning agent can be also developed if $q$ is unknown by adapting methods from~\cite{boyer} as described in~\cite{hamann}.

\subsubsection{Description of the agent}							

A hybrid agent that learns faster than a classical one can be constructed by alternating between quantum amplitude amplification and classical policy update by repeatedly performing the following steps:

\begin{enumerate}
\item Given the classical probability distribution $p(\vec{a})$,  determine the success probability $q$ based on the current policy and prepare the quantum state 
\begin{eqnarray}
\ket{\psi}=\sum\limits_{\lbrace \vec{a}\rbrace}\sqrt{p(\vec{a})}\ket{\vec{a}}.
\end{eqnarray}
\item Apply the optimal number of Grover iteration $k(\sqrt{q})$ leading to
\begin{eqnarray}
\centering
\ket{\psi'}=U_G^k\ket{\psi}
\end{eqnarray}
and perform a measurement on $\ket{\psi'}$ in the computational basis to determine a test sequence of actions $\vec{a}$. \\

\item Play one classical epoch by using the test sequence $\vec{a}$ determined in step 2 and record the corresponding sequence of percepts $\vec{s}(\vec{a})$ and the reward $r(\vec{a})$.
\item Update the classical policy $\pi(a_k|s_{k-1})$ and the resulting probability distribution $p(\vec{a})$ according to \ref{update}.
\end{enumerate}

There exists a limit $Q$ on $q$ determining whether it is more advantageous for the agent to perform a Grover iteration with $k(\sqrt{q})\geq 1$ or sample directly from $p(\vec{a})$ (therefore $k(\sqrt{q})=0$) to determine $\vec{a}$. In the latter case the agent would only interact classically (as in step 3) with the environment.

After each epoch, a classical agent receives a reward with probability $q = \varepsilon=\sin^2 (\xi)$. We assume that the agent can use one epoch to either perform one Grover iteration (step 2) or to determine the reward of a given test sequence $\vec{a}$ (step 3). Thus, for $k=1$ it receives after every second epoch a reward with probability $\sin^2(3\xi)$. As a consequence, we define the expected average reward of an agent playing a classical strategy as $\eta_C=2\sin^2(\xi)$ and of an agent playing a quantum strategy with $k=1$ as $\eta_Q=\sin^2 (3\xi)$. For $q<Q$, $\eta_Q>\eta_C$, meaning that the quantum strategy proves advantageous over the classical case. However, as soon as $\eta_C=\eta_Q$ (at $Q=0.396$), a classical agent starts outperforming a corresponding quantum agent which still performs Grover iterations.

Determining the winning probability $q$ exactly such that $q=\varepsilon$ as in the example presented here is not always possible. In general, additional information like the number of possible solutions and model building helps to perform this task. Note that a $Q$ smaller than $0.396$ should be chosen if $q$ can only be estimated up to some range.  To circumvent this problem, methods like Grover search with unknown reward probability~\cite{boyer}, or fixed-point search~\cite{yoder}, can be used to determine if and how many steps of amplitude amplification should be performed~\cite{hamann}.

\subsubsection{Learning time}		

We define the learning time $T$ as the number of epochs an agent needs on average to reach a predefined winning probability $Q_L$. The above described quantum-enhanced agent can reach the probability $Q$ on average with less epochs than its classical counterpart. However, once both reach $Q$, they need on average the same number of epochs to reach $Q+\Delta_Q$ with $0\leq\Delta_Q<1-Q$. Therefore, we choose $Q_L\leq Q$ in order to quantify the achievable improvement of a hybrid agent compared to its classical counterpart. In our experiment, we choose $Q_L=0.37$ to define the learning time.

Let $\ell_J=\lbrace \vec{a}_1,\cdots,\vec{a}_J\rbrace$ be a time-ordered list of all the rewarded sequences of actions an agent has found until it reaches $Q_L$.
Note that the  actual policy $\pi_j$, and thus $p_j$, of our agents depend only on the list $\ell_j$ of observed rewarded sequences of actions, and this is independent of whether it has found them via classical sampling or quantum amplitude amplification. As a result, a classical agent and its quantum-enhanced version are described by the same policy $\pi(\ell_j)$ and behave similar if they have found the same rewarded sequences of actions. However, the quantum-enhanced agent finds them faster. 

In general, the actual policy and overall success probability might depend on the found rewarded action sequence. Thus the number $J$ of observed rewarded action sequences necessary to learn might vary. However, this is not the case for the here reported experiment. The learning time in this case can be determined via
\begin{equation}
T(J)=\sum\limits_{j=1}^J t_j
\end{equation}   
where $t_j$ determines the number of epochs necessary for the agent to find the next rewarded  sequence $\vec{a}_j$ after it has observed $j-1$ rewards. For a purely classical agent, the average time is given by
\begin{equation}
\label{equ2}
\langle t_j\rangle_C=\frac{1}{q_j}
\end{equation}
where $q_j$ is the actual success probability. 
This time is quadratically reduced to 
\begin{equation}
\langle t_j\rangle_Q=\frac{\alpha}{\sqrt{q_j}}
\end{equation}
for the quantum-enhanced agent. Here, $\alpha$ is a parameter depending only on the number of epochs needed to create one oracle query $\tilde{U}_E$ \cite{dunjko3} and on whether $q_j$ is known. In the case considered here, we find $\alpha = \pi/4$. As a consequence, the average learning time for the quantum-enhanced learning agent is given by
\begin{eqnarray}
\langle T(J) \rangle_Q &=&  \sum\limits_{j=1}^J \frac{\alpha}{\sqrt{q_j}} \\
&\leq& \alpha \sqrt{J}\sqrt{\sum\limits_{j=1}^J \frac{1}{q_j}}\\
&\leq& \alpha \sqrt{J} \sqrt{\langle T(J)\rangle_C}
\end{eqnarray}    
where we used the Cauchy-Schwarz inequality in the second step. The classical learning time typically scales with $\langle T \rangle_C\sim A^K$ for a learning problem with episode length $K$ and the choice between $A$ different actions in each step. The number $J$ of observed rewarded sequences in order to learn depends on the specific policy update and sometimes also on the list $\ell_J$ of observed rewarded sequences of actions. For an agent sticking with the first rewarded action sequence, we would find $J=1$. However, typical learning agents are more explorative, and common scalings are $J\sim K$ such that we find for these cases
\begin{equation}
\langle T(J)\rangle_Q \sim \sqrt{\log(\langle T(J) \rangle_C)}\sqrt{\langle T(J) \rangle_C}.
\end{equation}
This is equivalent to a quasi-quadratic speed-up in the learning time if it is possible to perform arbitrary numbers of Grover iterations.

In more general settings, there exist several possible $\ell_J$ with different length $J$ such that the  learning time $\langle T (J)\rangle $ needs to be averaged over all possible $\ell_J$, which leads again to a quadratic speed-up in learning \cite{hamann}.

\subsubsection{Limited coherence times}

In general, all near-term quantum devices allow for coherent evolution only for a limited time and thus a maximal number $n$ of Grover iterations. For winning probabilities $q=\sin^2\xi$ with $(2n+1)\xi\leq \pi/2$, performing $n$ Grover iterations leads to the highest probability of finding a rewarded action.   

Again, we assume that the actual policy of an agent only depends on the number of observed rewards an agent has found. As a consequence, the average time a quantum-enhanced agent limited to $n$ Grover iterations needs to achieve the success probability $Q<\sin^2(\pi/(4n+2))$ is given by
\begin{equation}
\label{equ1}
\langle T(J,n)\rangle_Q = \sum\limits_{j=1}^J \frac{\alpha_0 n+1}{\sin[(2n+1)\xi_j]^2}
\end{equation} 

with $\sin^2 \xi_j= q_j$ and $\alpha_0$ determining the number of epochs necessary to create one oracle query $\tilde{U}_E$. For $\alpha_0=1$, $n>>1$ and $(2n+1)\xi_J\ll \pi/2$ we can approximate the learning time for the quantum-enhanced agent via
\begin{equation}
\langle T(J,n)\rangle_Q \approx \sum\limits_{j=1}^J \frac{1}{4nq_j}=\frac{\langle T(J)\rangle_C}{4n}
\end{equation}
where we used $\sin x\approx x$ for $x\ll 1$.
In general, it can be shown \cite{hamann} that the success probability $Q_n=\sin^2(\pi/(4n+2))$ can be reached by a quantum-enhanced agent limited to $n$ Grover iterations in a time
\begin{equation}
\langle T(n)\rangle_Q \leq \gamma \frac{\langle T\rangle_C}{n}
\end{equation}
where $\gamma$ is a factor depending on the specific setting. \\
In our case Eq. \ref{equ1} can be used to compute the lower bound for the average quantum learning time, with $\alpha_0=n=1$. For the classical strategy Eq. \ref{equ2} is used. 

\subsection{Experimental details}\label{Append2}

A continuous wave laser (Coherent Mira HP) is used to pump a single-photon source producing photon pairs in the telecom wavelength band. The laser light has a central wavelength of 789.5 nm and pumps the single photon source at a power of approximately 100 mW. The source is a periodically poled KTiOPO$_{4}$ non-linear crystal placed in a Sagnac interferometer~\cite{kim,saggio}, where the emission of single photons occurs via a type II Spontaneous Parametric Down-Conversion (SPDC) process. The crystal (produced by Raicol) is 30 mm long, set to a temperature of 25 $^{\circ}$C, has a poling period of 46.15 $\mu$m and is quasi-phase matched for degenerate emission of photons at 1570 nm when pumping with coherent laser light at 785 nm. As the processor is calibrated for a wavelength of 1580 nm, we shift the wavelength of the laser light to 789.75 nm in order to produce one photon at 1580 nm (that is then coupled into the processor) and another one at 1579 nm (the heralding photon).

The processor used for the experiment is a silicon-on-insulator (SOI) type, designed by the Quantum Photonics Laboratory at MIT (Massachusetts Institute of Technology)~\cite{harris}. 
Each programmable unit on the device acts as a tunable beam splitter implementing the unitary 
\begin{equation}
U_{\theta, \phi}=
\begin{pmatrix}
e^{i \phi} \sin \frac{\theta}{2}  &  e^{i \phi} \cos \frac{\theta}{2} \\
\cos \frac{\theta}{2} & - \sin \frac{\theta}{2}
\end{pmatrix}
\label{unitary}
\end{equation}
where $\theta$ and $\phi$ are the internal and external phases (as in Fig. \ref{chip}(b)) set via thermo-optical phase shifters controlled by a voltage supply. The achievable precision for phase settings is higher than $250$ $\mu$rad. The bandwidth of the phase shifters is around 130 kHz. The waveguides, spatially separated from one another by 25.4 $\mu$m, are designed to admit one linear polarization only. The high contrast in refractive index between the silicon and silica (the insulator) allows for waveguides with very small bend radius (less than 15 $\mu$m), thus enabling a high component density (in our case 88 MZIs) on small areas (in our case $4.9$ x $2.4$ mm). Given the small dimensions, the in- (and out-)coupling is realized with the help of $Si_3N_4-SiO_2$ waveguide arrays (produced by Lionix International), that shrink (and enlarge) the 10 $\mu$m  optical fibers' mode to match the 2 $\mu$m mode size of the waveguides in the processor. The total input-output loss is around 7 dB. The processor is stabilized to a temperature of 28 $^{\circ}$C and calibrated at 1580 nm for optimal performance. 
To reduce the black-body radiation emission due to the heating of the phase shifters when voltage is applied, wavelength division multiplexers with a transmission peak centered at 1571 nm and bandwidth of 13 nm are used before the photons are sent to the detection apparatus. In our processor, two external phase shifters in the implemented circuits were not responding to the supplied voltage. This defects were accounted for by deploying an optimization procedure. 

The single-photon detectors are multi-element superconducting nanowires (produced by photonSpot) with efficiencies up to 90\% in the telecom wavelength band. They have a dark count rate of $\backsim 100$ c.p.s, low timing jitter (hundreds of ps) and a reset time $<$ 100 ns~\cite{marsili}. 


\begin{thebibliography}{12}

\bibitem{sutton} Sutton, R. S. \& Barto, A. G., \textit{Reinforcement Learning: An Introduction} (MIT press, Cambridge, 1998).

\bibitem{johannink} Johannink, T. et al. Residual Reinforcement Learning for Robot Control. In \textit{2019 International Conference on Robotics and Automation (ICRA), Montreal, QC, Canada}, 6023-6029 (IEEE, 2019). URL https://doi.org/10.1109/ICRA.2019.8794127.

\bibitem{tjandra} Tjandra, A., Sakti, S. \& Nakamura, S. Sequence-to-Sequence ASR Optimization via Reinforcement Learning. In \textit{2018 IEEE International Conference on Acoustics, Speech and Signal Processing (ICASSP), Calgary, AB, Canada}, 5829-5833 (IEEE, 2018). URL https://doi.org/10.1109/ICASSP.2018.8461705.

\bibitem{komorowski} Komorowski, M., Celi, L. A., Badawi, O., Gordon, A. C. \& Faisal A. A. The artificial intelligence clinician learns optimal treatment strategies for sepsis in intensive care. \textit{Nat. Med.} \textbf{24,} 1716-1720 (2018). URL http://doi.org/10.1038/s41591-018-0213-5.

\bibitem{thakur} Thakur, C. S. et al. Large-scale neuromorphic spiking array processors: A quest to mimic the brain. \textit{Frontiers in neuroscience} \textbf{12,} 891 (2018). URL https://doi.org/10.3389/fnins.2018.00891. 

\bibitem{shen} Shen, Y. et al. Deep learning with coherent nanophotonic circuits. \textit{Nat. Photon.} \textbf{11,} 441 (2017). URL https://doi.org/10.1038/nphoton.2017.93. 

\bibitem{steinbrecher} Steinbrecher, G. R., Olson, J. P., Englund, D. \& Carolan, J. Quantum optical neural networks. \textit{npj Quantum Information} \textbf{5,} 1-9 (2019). URL https://doi.org/10.1038/s41534-019-0174-7. 

\bibitem{silver} Silver, D. et al. Mastering the game of Go without human knowledge. \textit{Nature} \textbf{550,} 354-359 (2017). URL http://dx.doi.org/10.1038/nature24270.

\bibitem{arute} Arute, F. et al. Quantum supremacy using a programmable superconducting processor. \textit{Nature} \textbf{574,} 505-510 (2019). URL https://doi.org/10.1038/s41586-019-1666-5.



\bibitem{giovannetti} Giovannetti, V., Lloyd, S. \& Maccone, L. Advances in quantum metrology. \textit{Nat. Photon.} \textbf{5,} 222 (2011). URL https://doi.org/10.1038/nphoton.2011.35. 

\bibitem{sergienko} Sergienko, A. V. (ed.) \textit{Quantum Communications and Cryptography} (CRC press, Boca Raton, 2018). URL https://doi.org/10.1201/9781315221120. 

\bibitem{monroe} Monroe, C. Quantum information processing with atoms and photons. \textit{Nature} \textbf{416,} 238-246 (2002). URL https://doi.org/10.1038/416238a. 

\bibitem{dong} Dong. D., Chen, C., Li, H \& Tarn, T.-J. Quantum Reinforcement Learning. \textit{IEEE T. Syst, Man Cy. B} \textbf{38,} 1207-1220 (2008). URL https://doi.org/10.1109/TSMCB.2008.925743.

\bibitem{dunjko} Dunjko, V. \& Briegel, H. J. Machine learning \& artificial intelligence in the quantum domain: a review of recent progress. \textit{Rep. Progr. Phys.} \textbf{81,} 074001 (2018). URL https://doi.org/10.1088/1361-6633/aab406.

\bibitem{baireuther} Baireuther, P., O'Brien, T. E., Tarasinski, B. \& Beenakker, C. W. J. Machine-learning-assisted correction of correlated qubit errors in a topological code. \textit{Quantum} \textbf{2,} 48 (2018). URL https://doi.org/10.22331/q-2018-01-29-48. 

\bibitem{breuckmann} Breuckmann, N. P. \& Ni, X. Scalable Neural Network Decoders for Higher Dimensional Quantum Codes. \textit{Quantum} \textbf{2,} 68-92 (2018). URL https://doi.org/10.22331/q-2018-05-24-68. 

\bibitem{chamberland} Chamberland, C. \& Ronagh, P. Deep neural decoders for near term fault-tolerant experiments. \textit{Quant. Sci. Techn.} \textbf{3,} 044002 (2018). URL https://doi.org/10.1088/2058-9565/aad1f7.  

\bibitem{fosel} F\"{o}sel, T., Tighineanu, P., Weiss, T. \& Marquardt, F. Reinforcement Learning with Neural Networks for Quantum Feedback. \textit{Phys. Rev. X} \textbf{8,} 031084 (2018). URL https://doi.org/10.1103.PhysRevX.8.031084. 

\bibitem{poulsen} Poulsen Nautrup, H., Delfosse, N., Dunjko, V., Briegel, H. J. \& Friis, N. Optimizing quantum error correction codes with reinforcement learning. \textit{Quantum} \textbf{3,} 215 (2019). URL https://doi.org/10.22331/q-2019-12-16-215.

\bibitem{krenn} Krenn, M., Malik, M., Fickler, R., Lapkiewicz, R. \& Zeilinger, A. Automated Search for new Quantum Experiments. \textit{Phys. Rev. Lett.} \textbf{116,} 090405 (2016). URL http://doi.org/10.1103/PhysRevLett.116.090405.

\bibitem{melnikov} Melnikov, A. A. et al. Active learning machine learns to create new quantum experiments. \textit{Proc. Natl. Acad. Sci. U.S.A.} \textbf{115,} 1221-1226 (2018). URL http://dx.doi.org/10.1073/pnas.1714936115. 

\bibitem{paparo} Paparo, G. D., Dunjiko, V., Makmal, A., Martin-Delgrado, M. A. \& Briegel, H. J. Quantum Speedup for Active Learning Agents. \textit{Phys. Rev. X} \textbf{4,} 031002 (2014). URL http://dx.doi.org/10.1103/PhysRevX.4.031002.

\bibitem{dunjko2} Dunjko, V., Friis, N. \& Briegel, H. J. Quantum-enhanced deliberation of learning agents using trapped ions. \textit{New J. Phys.} \textbf{17,} 023006 (2015). URL https://doi.org/10.1088/1367-2630/17/2/023006. 

\bibitem{jerbi} Jerbi, S., Poulsen Nautrup, H., Trenkwalder, L. M., Briegel, H. J. \& Dunjko, V. A framework for deep energy-based reinforcement learning with quantum speed-up (2019). Preprint at https://arxiv.org/abs/1910.12760.  

\bibitem{sriarunothai} Sriarunothai, T. et al. Speeding-up the decision making of a learning agent using an ion trap quantum processor. \textit{Quantum Sci. Technol.} \textbf{4,} 015014 (2019). URL http://dx.doi.org/10.1088/2058-9565/aaef5e.

\bibitem{dunjko3} Dunjko, V., Taylor, J. M., \& Briegel, H. J. Quantum-Enhanced Machine Learning. \textit{Phys. Rev. Lett.} \textbf{117,} 130501 (2016). URL http://dx.doi.org/10.1103/PhysRevLett.117.130501.

\bibitem{kimble} Kimble, H. J. The quantum internet. \textit{Nature} \textbf{453,} 1023-1030 (2008). URL https://doi.org/10.1038/nature07127. 

\bibitem{cacciapuoti} Cacciapuoti, A. S. et al. Quantum Internet: Networking Challenges in Distributed Quantum Computing. \textit{IEEE Network} \textbf{34,} 137-143 (2020). URL https://doi.org/10.1109/MNET.001.1900092. 


\bibitem{denil} Denil, M. et al. Learning to Perform Physics Experiments via Deep Reinforcement Learning (2016). Preprint at https://arxiv.org/abs/1611.01843. 

\bibitem{bukov} Bukov, M. et al. Reinforcement Learning in Different Phases of Quantum Control. \textit{Phys. Rev. X} \textbf{8,} 031086 (2018). URL http://dx.doi.org/10.1103/PhysRevX.8.031086. 

\bibitem{poulsen2} Poulsen Nautrup, H. et al. Operationally meaningful representations of physical systems in neural networks (2020). Preprint at https://arxiv.org/abs/2001.00593.

\bibitem{grover} Grover, L. K. Quantum mechanics helps in searching for a needle in a haystack. \textit{Phys. Rev. Lett.} \textbf{79,} 325 (1997). URL https://doi.org/10.1103/PhysRevLett.79.325. 
 
\bibitem{flamini} Flamini, F. et al. Photonic architecture for reinforcement learning. \textit{New. J. Phys.} \textbf{22,} 045002 (2020). URL https://doi.org/10.1088/1367-2630/ab783c. 

\bibitem{yoder} Yoder, T. J., Low, G. H. \& Chuang, I. L. Fixed-Point Quantum Search with an Optimal Number of Queries. \textit{Phys. Rev. Lett.} \textbf{113,} 210501 (2014). URL https://doi.org/10.1103/PhysRevLett.113.210501. 

\bibitem{hamann} Hamann, A. et al. A hybrid agent for quantum-accessible reinforcement learning (2020). In preparation.

\bibitem{boyer} Boyer, M., Brassard, G., Hoyer, P. \& Tappa, A. Tight bounds on quantum searching. \textit{Fortschr. Phys.} \textbf{46,} 493 (1998). URL https://doi.org/10.1002/3527603093.ch10. 

\bibitem{briegel} Briegel, H. J. \& De las Cuevas, G. Projective simulation for artificial intelligence. \textit{Sci. Rep.} \textbf{2,} 1-16 (2012). URL http://doi.org/10.1038/srep00400.

\bibitem{kim} Kim, T., Fiorentino, M. \& Wong, F. N. C. Phase-stable source of polarization-entangled photons using a polarization Sagnac interferometer. \textit{Phys. Rev. A} \textbf{73,} 012316 (2006). URL http://doi.org/10.1103/PhysRevA.73.012316. 

\bibitem{saggio} Saggio, V. et al. Experimental few-copy multipartite entanglement detection. \textit{Nat. Phys.} \textbf{15,} 935-940 (2019). URL http://doi.org/10.1038/s41567-019-0550-4. 

\bibitem{harris} Harris, N. C. et al. Quantum transport simulations in a programmable nanophotonic processor. \textit{Nat. Photon.} \textbf{11,} 447-452 (2017). URL http://doi.org/10.1038/nphoton.2017.95. 

\bibitem{marsili} Marsili, F. et al. Detecting single infrared photons with 93\% system efficiency. \textit{Nat. photon.} \textbf{7,} 210-214 (2013). URL https://doi.org/10.1038/nphotonc.2013.13.



\end{thebibliography}
\end{document}